\documentstyle[12pt]{article}
\newcommand{\Frac}[2]{\frac{\displaystyle #1}{\displaystyle #2}}
\newcommand{\cu}{^{\circ}}

\newcommand{\ggpp}{\gamma \gamma \rightarrow \pi \pi}
\newcommand{\ggppcero}{\gamma \gamma \rightarrow \pi^{\circ} \pi^{\circ}}
\newcommand{\ggppplus}{\gamma \gamma \rightarrow \pi^+ \pi^-}
\newcommand{\gpgp}{\gamma \pi \rightarrow \gamma \pi}
\newcommand{\gpgpcero}{\gamma \pi^{\circ} \rightarrow \gamma \pi^{\circ}}
\newcommand{\gpgpplus}{\gamma \pi^+ \rightarrow \gamma \pi^+}
\newcommand{\alphap}{\alpha_{\pi}}
\newcommand{\betap}{\beta_{\pi}}
\newcommand{\alphaplus}{\alpha_{\pi^{\pm}}}

\newcommand{\alphapcero}{\alpha_{\pi^{\circ}}}
\newcommand{\betapcero}{\beta_{\pi^{\circ}}}
\newcommand{\aplusbc}{(\alpha_{\pi} + \beta_{\pi})^C}
\newcommand{\aminusbc}{(\alpha_{\pi} - \beta_{\pi})^C}
\newcommand{\aplusbn}{(\alpha_{\pi} + \beta_{\pi})^N}
\newcommand{\aminusbn}{(\alpha_{\pi} - \beta_{\pi})^N}
\begin{document}
\pagestyle{empty}
\begin{titlepage}
\begin{center}
\hfill DTP-94/52 \\
\hfill July, 1994 \\
\vspace*{4cm}
{\LARGE \bf  Theoretical Predictions}
\vspace*{1cm} \\
{\LARGE \bf for}
\vspace*{1cm} \\
{\LARGE \bf Pion Polarizabilities \footnote{To be published in the
Second DA$\Phi$NE Physics Handbook, ed. G. Pancheri and N. Paver.}}
\vspace*{2.5cm} \\
{\Large  J. Portol\'es, $\, \,$   M.R. Pennington}  \\
\vspace*{0.5cm}
Centre for Particle Theory,\\ University of Durham, \\
Durham, DH1 3LE,  (U.K.)
\vspace*{3.5cm} \\
\begin{abstract}
\baselineskip=6mm
\noindent The polarizabilities of the pion have been predicted in several
different theoretical frameworks. The status of these is
reviewed.
\end{abstract}
\end{center}
\end{titlepage}
\newpage
\pagestyle{plain}
\pagenumbering{arabic}
\baselineskip=6.8mm
\parskip=1.5mm
\section{Introduction}

\hspace{0.5cm}
The concept of polarizability first appeared  in the realm of particle
physics twenty years ago \cite{TE73,TE74,FR75,BT76} as a quantity
which characterizes an
elementary particle, like its charge radius, magnetic moment, etc.
In {\it classical} physics the polarizability of a medium (or a composite
system in general) is a well known concept related to the response of
the system to the presence of an external electromagnetic field,
representing a measure of how easy it is to polarize it.
The translation of this quantity into {\it quantum} physics involves
 Compton scattering on the corresponding target. For an electrically
charged system, scattering at threshold is determined by the charge
of the system. This is the Thompson limit. The polarizabilities give the
corrections to Thompson scattering  ---corrections to the next order in the
energy of the photons. For neutral targets, the corrections parameterized
by the polarizabilities are the leading answer.
\par
 Since the introduction of this new quantity, considerable
 work has been performed both theoretically and experimentally,
 largely on the nucleon and pion polarizabilities.
 In this article we focus attention on the theoretical
predictions for the pion polarizabilities.
\par
All the numerical results about polarizabilities given in this
paper are expressed in the Gaussian system ($e^2 = \alpha$) and in
units of $10^{-43} \rm{cm^3} = 10^{-4} \rm{fm^3}$, which are not
quoted.

\section{Definition of pion polarizabilities}

\hspace{0.5cm} Let us consider Compton scattering on a pion
\begin{equation}
\gamma (q_1) \, \pi (p_1) \longrightarrow \gamma (q_2) \, \pi (p_2)
\end{equation}
where $q_i$ are the 4--momenta of the photons and $p_i$ those for the
pions. The amplitude for this process can then be expanded in powers
of the energies of the photons:
\begin{equation}
A(\gpgp)|_{threshold} \;  =  \; \left[ \, - \Frac{\alpha}{m_{\pi}} \delta_{\pi
\pi^{\pm}} + \alphap \, \omega_1 \omega_2 \, \right] \, \hat{\epsilon}_1
\cdot \hat{\epsilon}_2^*
 + \betap \, \omega_1 \omega_2 \, ( \hat{\epsilon}_1 \times
\hat{q}_1 ) \cdot ( \hat{\epsilon}_2^* \times \hat{q}_2 ) + \ldots
\label{eq:def}
\end{equation}
where  $q_i = \omega_i \, ( 1 , \hat{q_i})$ for $i = 1,2$ and
$\hat{\epsilon}_i$ is the polarization vector of the photon with momentum
$q_i$.
\par
In Eq. (\ref{eq:def}) $\alphap$ and $\betap$ are the electric and magnetic
polarizabilities, respectively, and $\alpha$ is the fine structure
constant. Let us note:
\begin{itemize}
\item[1.-] At zeroth order in the energies of the photons only the
{\it point-like} structure of the target survives and it is accordingly
zero for a neutral pion.
\item[2.-] There is no linear term in the energy of photons. This is
because we are dealing  with a spinless target. In the case of
the nucleon, for example, there is a non--vanishing linear term.
\item[3.-] The polarizabilities are not specified by symmetry arguments
alone.
\end{itemize}
As can be seen from their definition, the polarizabilities carry
information about the electromagnetic structure of the target in the
Compton process.

\section{General remarks on pion polarizabilities}

\hspace{0.5cm} Before detailing various theoretical
predictions, there are results we know about the
pion polarizabilities on quite general grounds:

\begin{itemize}

\item[1/] Chiral dynamics demands that in the exact chiral limit the
relation
\begin{equation}
\alpha_{\pi} + \beta_{\pi} = 0
\label{eq:suma}
\end{equation}
must be obeyed \cite{DH89}.

\item[2/] A dispersion relation for the forward scattering
amplitude gives \cite{TE74,HO90}
\begin{equation}
\alpha_{\pi} + \beta_{\pi} = \Frac{1}{2 \pi^2} \int_0^{\infty}
d \omega \, \, \, \Frac{\sigma_{tot} (\omega)}{\omega^2} \, \, \, \, \, \, \,
\, \, ,
\end{equation}
where $\sigma_{tot}$ is the total photoproduction cross section
on pions; this, of course, implies
\begin{equation}
\alpha_{\pi} + \beta_{\pi} > 0 \; \; \; \; \; \; \; .
\end{equation}

\item[3/]  Crossing  symmetry relates the polarizabilities  to  the
helicity amplitudes of the process $\ggpp$ at the crossed--channel
threshold. If we call $M_{++}$ the helicity 0 amplitude and $M_{+-}$ the
helicity 2  amplitude \cite{KS86,BG93} for $\ggpp$,
\begin{eqnarray}
(\alpha_{\pi} \pm \beta_{\pi})^C & = &
- \Frac{\alpha}{m_{\pi}} \, \left[ \, M_{+ \mp}^C - M_{BORN} \, \right] \,
\Big|_{s=0, t= m_{\pi}^2} \nonumber \\
\label{eq:sumdif}
& & \\
(\alpha_{\pi} \pm \beta_{\pi})^N & = &
\Frac{\alpha}{m_{\pi}} \, M_{+ \mp}^N \, |_{s=0, t= m_{\pi}^2} \, \, \,
\, \, \, \, \, ,
\nonumber
\end{eqnarray}
where the superscript $C$ or $N$ denotes $charged$ or $neutral$
pions, respectively. In the charged case, the Born
amplitude must be subtracted first to obtain the corresponding
combination of electric and magnetic polarizabilities. The
relations, Eq. (\ref{eq:sumdif}), allow us to see that the combination
$(\alpha_{\pi} - \beta_{\pi})$ is pure S--wave, while $(\alpha_{\pi} +
\beta_{\pi})$ is pure D--wave in the $\ggpp$ channel.

\item[4/] The polarizabilities of the charged pion are directly related by
chiral dynamics to the axial ($h_A$) and vector ($h_V$)
structure--dependent form factors of the radiative pion decay
$\pi^+ \rightarrow e^+ \nu_e \gamma$ as \cite{DH89}
\begin{equation}
\alphaplus = \Frac{\alpha}{8 \pi^2 m_{\pi} F_{\pi}^2} \Frac{h_A}{h_V}
\label{eq:hav}
\end{equation}
where $F_{\pi} \sim 93 \, \rm{MeV}$ is the decay constant of pion.

\item[5/] A sum rule for the electric polarizability was proposed by
Petrun'kin
\cite{PE64} (see also \cite{EH73,BH88}) using a classical approach.
This sum rule says that $\alphap$ can be split into two parts
\begin{equation}
\alphap = \alphap^{cl} + \alphap^{intr}  \, \, \, \, \, \, \, \, \, .
\end{equation}
The term $\alphap^{cl}$ is related to the electromagnetic pion size
(it is proportional to the charge radius squared)
\begin{equation}
\alphap^{cl} = \Frac{\alpha}{3 m_{\pi}} \langle r_{\pi}^2 \rangle
\, \, \, \, \, \, \, \, \, ,
\end{equation}
while $\alphap^{intr}$ is the intrinsic polarizability associated
with possible excited states of the pion accessed by electric dipole
transitions and vacuum polarization effects:
\begin{equation}
\alphap^{intr} = 2 \alpha \sum_{n \neq 0}
\Frac{| \langle n | {\cal{D}} | 0 \rangle |^2}{E_n - E_{\circ}} \,
\, \, \, \, \, \, \, \, ,
\label{eq:intr}
\end{equation}
where $\cal{D}$ is the electric dipole operator.
This description has been criticized by Terent'ev \cite{TE74}
(on grounds that assumptions involved in taking the
non--relativistic limit are dubious), but Holstein \cite{HO90} gives an
interpretation of $\alphap^{intr}$  relating
it to the spectral functions of vector and axial--vector mesons,
$\rho^V (s)$ and $\rho^A (s)$, respectively, by comparing with the
known
current algebra result \cite{DM67}. This gives,
\begin{equation}
\sum_{n \neq 0}
\Frac{| \langle n | {\cal{D}} | 0 \rangle |^2}{E_n - E_{\circ}} =
\, \Frac{1}{4 m_{\pi} F_{\pi}^2}
\int d s \, \, \Frac{\rho^A (s) - \rho^V (s)}{s^2}
\label{eq:espectro}
\end{equation}
Bernard et al. \cite{BH88}
have computed the intrinsic electric polarizability, Eq. (\ref{eq:intr}),
and the analogous intrinsic magnetic polarizability in a valence
quark model (MIT bag). However, they obtain the opposite sign for the
magnetic polarizability compared to that given by phenomenology.
This result the
authors claim means that the pion cannot be regarded as a single
bound $q \, \overline{q}$
valence pair, but rather the pion must be treated as a
collective degree of
freedom to obtain consistent results.
\end{itemize}

\section{Experimental situation}

\hspace{0.5cm}It is evident that experimentally Compton scattering on
 pions is not
easy. Fortunately,  there are  processes,
when properly analyzed, that allow  information about pion
polarizabilities to be extracted.
In this section we comment briefly on
the status of these experimental determinations and
refer the reader to the reference \cite{BB92} for a more complete
discussion.
The experimental situation is very different for charged and
neutral pions:
\vspace*{0.5cm} \\
{\it Charged pion polarizabilities}
\vspace*{0.4cm} \\
All the experimental results (except when otherwise stated) are
analysed assuming the constraint on the sum of polarizabilities,
Eq. (\ref{eq:suma}).
The experimental sources of information and results are:
\begin{itemize}
\item[1/] {\em Radiative pion nucleon scattering (Primakoff effect)}
$[ \, \pi^- Z \rightarrow \pi^- Z \gamma \, ]$
\newline
  The SERPUKOV group
gives \cite{AN83}
\begin{equation}
\alphaplus= 6.8 \pm 1.4 (stat) \pm 1.2 (syst) \; \; \; \; \; \; \; .
\label{eq:serpukov}
\end{equation}
When they relax the condition Eq. (\ref{eq:suma}), they find \cite{AN85}
\begin{equation}
\aplusbc = 1.4 \pm 3.1 (stat) \pm 2.5 (syst) \; \; \; \; \; \; \; ,
\end{equation}
which is consistent with Eq. (\ref{eq:suma}).

\item[2/] {\em Pion photoproduction in photon--nucleon scattering}
$[ \, \gamma p \rightarrow \gamma \pi^+ n\, ]$
\newline
 The LEBEDEV group
obtains \cite{AI86}
\begin{equation}
\alphaplus = 20 \pm 12 (stat) \; \; \; \; \; \; \; .
\end{equation}

\item[3/] {\em Photon--photon into two pions} $[ \, \ggppplus \, ]$
\newline
The data of the MARK II group \cite{BO90} have been analyzed in
\cite{BB92} with the result
\begin{equation}
\alphaplus= 2.2 \pm 1.6 (stat + syst) \; \; \; \; \; \; \; .
\end{equation}
\end{itemize}
As can be seen  the results are far from consistent with each other. Clearly
 more experimental effort is needed.
\vfil\eject
\noindent {\it Neutral pion polarizabilities}
\vspace*{0.4cm} \\
There are no real experimental measurements of neutral pion polarizabilities.
However,
Babusci et al.  \cite{BB92} have analyzed
data  on $\ggppcero$ taken by the Crystal Ball Collaboration \cite{MA90}
using the theoretical
calculation in Chiral Perturbation Theory to be discussed in
Sect. 5.1 (with the constraint
on the sum of polarizabilities Eq. (\ref{eq:suma})) and find,
\begin{equation}
| \alphapcero | = 0.69 \pm 0.07 (stat) \pm 0.04 (syst) \; \; \; \; \; \; \; .
\end{equation}
Other analyses involve parameterizations using dispersion relations
\cite{KS86,KS92} and then fitting the data. We will comment on these
in Sect. 5.4.

\section{Theoretical predictions}
\hspace*{0.5cm}
In spite of the fact that the concept of the polarizability of an elementary
particle was introduced long ago \cite{PE64,TE73,TE74}, only after the
first measurement of charged pion polarizabilities was this issue taken up
by  theorists. In the last two years there has been a burst
of theoretical predictions on charged and neutral pion polarizabilities.
These we collect here. We detail
 the theoretical frameworks employed, their respective
results and give a brief analysis of these.

\subsection{Chiral Perturbation Theory ($\chi PT$)}
\hspace*{0.4cm}
The study of the cross--section for the process $\ggpp$ in $\chi PT$
gave the first predictions. The leading contribution is ${\cal{O}}(p^4)$
in the chiral expansion and was computed by Bijnens and Cornet
\cite{BC88} using the $SU(3)_L \otimes SU(3)_R$ chiral Lagrangian
\cite{GL85}. At this order, even with $m_{\pi} \neq 0$, they found that
\begin{equation}
(\alpha + \beta)^{C,N} |_{\chi PT [{\cal{O}}(p^4)]} = 0
\end{equation}
and the results
\begin{eqnarray}
\alphaplus & = & \Frac{4 \alpha}{m_{\pi} F_{\pi}^2} \, ( \; L_9^r +
L_{10}^r \;) = 2.68 \pm 0.42
\nonumber \\
\label{eq:alp4}
& & \\
\alphapcero & = & - \Frac{\alpha}{96 \pi^2 m_{\pi} F_{\pi}^2} = -0.50
\; \; \; \; \; \; \; ,
\nonumber
\end{eqnarray}
where the error in $\alphaplus$ comes from the phenomenological determination
of  $L_9^r$ and $L_{10}^r$. It is worth emphasizing that there is no
contribution from the 1--loop graphs to $\alphaplus$ and that only the
pion loop contributes to $\alphapcero$.

$$
\begin{array}{|c|c|cc|c|}
\multicolumn{5}{c}{{\bf Table \; \; I} \; \; Predictions \; \; of \; \;
\chi P T \; \; for \; \; pion \; \; polarizabilities}
\vspace{0.5cm} \\
\hline
\hline
& & & & \\
\multicolumn{1}{|c}{Polarizability} &
\multicolumn{1}{|c}{{\cal{O}}(p^4)} &
\multicolumn{2}{|c}{{\cal{O}}(p^6) \; \cite{BG93}} &
\multicolumn{1}{|c|}{Total} \\
\cline{3-4}
& \cite{BC88} &
\multicolumn{1}{|c}{tree} &
\multicolumn{1}{|c|}{two-loop} & \\
\hline
\hline
& & & & \\
\alpha_{\pi^{\circ}} & -0.50 & 0.21 & -0.07 & -0.35 \pm 0.10 \\
& & & & \\
\hline
& & & & \\
\beta_{\pi^{\circ}}  &  0.50 & 0.79 & 0.24 & 1.50 \pm 0.20 \\
& & & & \\
\hline
& & & & \\
\aplusbn & 0 & 1.00 & 0.17 & 1.15 \pm 0.30 \\
& & & & \\
\hline
& & & & \\
\aminusbn & -1.00 & -0.58 & -0.31 & -1.90 \pm 0.20 \\
& & & & \\
\hline
\hline
& & & & \\
\alpha_{\pi^{\pm}} & 2.68 & - & - & 2.68 \pm 0.42 \\
& & & & \\
\hline
\hline
\end{array}
$$
\vspace {6mm}

Recently the next--to--leading order corrections to $\ggppcero$ in
$SU(2)_L \otimes
SU(2)_R$ $\chi PT$ have been calculated by Bellucci et al. \cite{BG93}.
The authors
 have computed the ${\cal{O}}(p^6)$ contribution that involves
a full two--loop calculation \footnote{See S. Bellucci in this Handbook.}.
In order to handle the divergences, the ${\cal{L}}^6$ contact terms of the
effective Lagrangian must be included. The strengths of these terms are
given by coupling constants, the values of which are
extracted by saturating with vector mesons $(1^{--})$, C--odd axial vector
mesons $(1^{+-})$, scalars $(0^{++})$ and tensors $(2^{++})$.
\par
At this order Eq. (\ref{eq:suma}) is no longer satisfied. Now
\begin{equation}
(\alpha + \beta)^{C,N}_{\chi P T [{\cal{O}}(p^6)]} \neq 0 \; \; \; \; \; \;
\; .
\end{equation}
The complete results are shown in Table I. The errors in the total
predictions come from the uncertainties in the phenomenological
determination of the coupling constants in the chiral Lagrangian.
\newpage
\subsection{Lowest order $\chi$PT with explicit resonance contributions}
\hspace*{0.4cm}
Another way to take into account corrections to the leading order result
comes from explicitly computing the resonance contributions. Chiral power
counting establishes that only axial vector mesons $(1^{++})$ start to
contribute to ${\cal{O}}(p^4)$, while vectors $(1^{--})$, C--odd axial
vectors $(1^{+-})$, scalars $(0^{++})$ and tensors $(2^{++})$ start at
${\cal{O}}(p^6)$. However such calculations do mean that part of the
higher order corrections
are automatically included in these resonance terms.

As the leading order is ${\cal{O}}(p^4)$ and in order to avoid double
counting the contribution of axial vectors $a_1 (1^{++})$ is not included.
This is because these resonances only contribute to $\alphaplus$, but the
combination $L_9^r + L_{10}^r$ in Eq. (\ref{eq:alp4}) comes from pure $a_1$
annihilation in the s--channel for $\gpgpplus$ \cite{DH93}
\footnote{This is so when
the Weinberg relation of masses $m_{a_1} = \sqrt{2} m_{\rho}$ is used.}.
We now enumerate the contribution of these resonances in turn:
\vspace*{0.4cm} \\
{\em Vector mesons $(1^{--}) \; [ \, \rho , \omega \, ]$}
\vspace*{0.4cm} \\
These resonances only modify the value of $\betap$. The contribution of their
direct channel exchange in  $\gpgp$ has been calculated in \cite{KO90,BB93} and
found to be:
\begin{eqnarray}
(\alpha_{\pi} + \beta_{\pi}) \, |^C_{\rho} & = & 0.07 \nonumber \\
\label{eq:vector}
& & \\
(\alpha_{\pi} + \beta_{\pi}) \, |^N_{\rho,\omega} & = & 0.83 \, \, \, \, \, \,
\, \, \, \, .\nonumber
\end{eqnarray}
Since the couplings of the light vector mesons are well--known , the
calculation of their contributions should be reliable.
\vspace*{0.4cm} \\
{\em C--odd axial vector  mesons $(1^{+-}) \; [ \, b_1 , h_1  \, ]$}
\vspace*{0.4cm} \\
 Ko has worked out
\cite{KO93} their contribution to $\gpgpcero$, and
hence  to the neutral pion polarizabilities. These resonances do not
modify $\betapcero$ and Ko finds \footnote{A numerical error
has been corrected here (we thank S. Bellucci for pointing this out).}
\begin{equation}
\alphapcero |_{b_1,h_1} = 0.21
\label{eq:codd}
\end{equation}
to be added to
\begin{equation}
\alphapcero |_{\chi P T [ {\cal{O}}(p^4) ]} = -0.50
\, \, \, \, \, \, \, \, \, .
\end{equation}
We note that since this resonance sector is not so well known,
 the result in Eq. (\ref{eq:codd}), which is  obtained assuming no mixing and
exact nonet symmetry, could change when more realistic approximations are
made. \footnote{By the way this warning must be extended also to the
${\cal{O}}(p^6)$ calculation in $\chi P T$ \cite{BG93} that
 model the C--odd axial vector mesons in the same way.}
\vspace*{0.4cm} \\
{\em Scalar mesons $(0^{++}) \; [ \, a_{\circ} , f_{\circ} \, ]$ and
tensor mesons $(2^{++}) \; [ \, a_2 , f_2 \, ]$}
\vspace*{0.4cm} \\
After \cite{AB92}, Babusci et al. \cite{BB93} have considered the
contribution of scalar and tensor resonances  to $\gpgpcero$. However
 a problem arises here. Since exchanges contribute in the
t--channel, an ambiguity results in the phenomenological determination
of the signs of the couplings. The contribution given by
tensor exchange is thus
\begin{equation}
| \alphapcero |_T = 0.04 \; \; \; \; \; \; \; , \; \; \; \; \; \; \;
| \betapcero |_T = 0.07 \; \; \; \; \; \; \; .
\end{equation}
The contribution from scalars follows from \cite{BG93} as
\begin{equation}
| \alphapcero |_S = 0.02 \; \; \; \; \; \; \; , \; \; \; \; \; \; \;
| \betapcero |_S = 0.02
\end{equation}
with $(\alpha_{\pi}^S + \beta_{\pi}^S)^N = 0$.
Since we only know the magnitudes, both these contributions  must be included
 as an uncertainty in the chiral prediction.
\par
In Table II
we collect the different resonance structure contributions. It is worth
 emphasizing that the results in Table II are not directly comparable
with those in Table I because of differences in the order of chiral powers
included.

$$
\begin{array}{|c|c|c|c|c|c|c|}
\multicolumn{7}{c}{{\bf Table \; \; II} \; \; Lowest \; \; order \; \;
\chi PT \; \;  with \; \;  explicit \; \; resonance \; \; contributions} \\
\multicolumn{7}{l}{\; \; \; \; \; \; \; \; \; \; \; \; \; \; \; \; \;
to \; \; pion \; \; polarizabilities}
\vspace{0.5cm} \\
\hline
\hline
& & & & & & \\
\multicolumn{1}{|c}{Polarizability} &
\multicolumn{1}{|c}{\chi P T |_{{\cal{O}}(p^4)}} &
\multicolumn{1}{|c}{1^{--}} &
\multicolumn{1}{|c}{1^{+-}} &
\multicolumn{1}{|c}{0^{++}} &
\multicolumn{1}{|c}{2^{++}} &
\multicolumn{1}{|c|}{Total} \\
& & & & & & \\
\hline
\hline
& & & & & & \\
\alpha_{\pi^{\circ}} & -0.50 & 0  & 0.21 & \pm 0.02 & \pm 0.04 & -0.29
\pm 0.04 \\
& & & & & & \\
\hline
& & & & & & \\
\beta_{\pi^{\circ}}  &  0.50 & 0.83 & 0 & \mp 0.02 & \pm 0.07 & 1.33
\pm 0.07 \\
& & & & & & \\
\hline
\hline
& & & & & & \\
\alpha_{\pi^{\pm}}  & 2.68 \pm 0.42 & 0 & -^*  & -^* & -^* & 2.68 \pm 0.42 \\
& & & & & & \\
\hline
& & & & & & \\
\beta_{\pi^{\pm}} & -2.68 \pm 0.42 & 0.07 & 0  & -^* & -^* & -2.61 \pm 0.42 \\
& & & & & & \\
\hline
\hline
\multicolumn{7}{l}{(*) This \; \; contribution \; \; has \; \; not \; \;
been \; \; calculated}  \\
\end{array}
$$

\subsection{Generalized Chiral Perturbation Theory $(G \chi P T)$}
\hspace*{0.5cm}
The cross--section of $\gpgpcero$ has recently been studied in the framework
of $G \chi P T$ \cite{SS93} by Knecht et al. \cite{KM94}.
This has allowed predictions to ${\cal{O}}(p^5)$ for the
neutral pion polarizabilities to be calculated. The authors have also
worked out  an
${\cal{O}}(p^4)$ result for the  charged pion polarizability. To these orders,
they still find
\begin{equation}
(\alpha_{\pi} + \beta_{\pi} )^{C,N} = 0 \, \, \, \, \, \, \, \, \, .
\end{equation}
The neutral pion polarizability  depends on 5 unknown parameters :
$\alpha_{\pi \pi} , \beta_{\pi \pi}$ (related to ${\cal{O}}(p^3)$
tree level $\pi \pi$ scattering), $\alpha_{\pi K} , \beta_{\pi K}$
(related to ${\cal{O}}(p^3)$ tree level $\pi K$ scattering) and a
combination of the coupling constants of the ${\cal{L}}^5_{G \chi P T}$
contact term they call $c$. Disregarding the kaon contribution
 (i.e. setting $\alpha_{\pi K} = \beta_{\pi K} = 0$), taking values for
$\alpha_{\pi \pi} ,
\beta_{\pi \pi}$ from a fit to  $\pi \pi$ scattering data \cite{SS93}
and constructing a low energy sum rule for $c$ evaluated by saturating
with the vector resonances, $\rho, \omega $ and $\phi$, they obtain
\begin{equation}
\alphaplus |_{{\cal{O}}(p^4)} = 3.47 \; \; \; \; \; \; \; \; \; \; \;
, \; \; \; \; \; \alphapcero |_{{\cal{O}}(p^5)} = + 0.44 \, \, \, \,
\, \, \, \, \, .
\end{equation}
Let us  note that the prediction to ${\cal{O}}(p^5)$ of $G \chi P T$
for $\alphapcero$ has a different sign than the prediction to
${\cal{O}}(p^6)$ in $\chi P T$ (Table I).

\subsection{Dispersive descriptions}
\hspace*{0.5cm}
\baselineskip=7.2mm
The S--wave isospin amplitudes for the processes $\ggpp$ were worked out
by  Morgan and  Pennington \cite{MP91} using twice--subtracted
dispersion relations
\begin{equation}
f_I(s) = \Omega_I(s) \, \Bigg\{ \, p_I(s) \Omega_I^{-1}(s) + c_I + s d_I  -
\Frac{(s-s_{\circ})^2}{\pi} \int_{4 m_{\pi}^2}^{\infty} dx
\Frac{p_I(x) \Im m [ \Omega_I^{-1}(x) ]}{(x-s_{\circ})^2 (x - s - i \epsilon)}
 \Bigg\}
\label{eq:disp}
\end{equation}
for $I=0,2$, where $\Omega_I(s)$ is the Omn\`es function, $p_I(s)$ gives the
structure of the left hand cut, and $c_I, d_I$ are two subtraction
constants which depend on the subtraction point $s_{\circ}$. Analogous
relations hold
for higher waves, their known threshold behaviour fixing their subtraction
constants \cite{MP88}.
\par
Once the $f_I(s)$ are determined it is possible to deduce the combination
$(\alpha_{\pi} - \beta_{\pi})$ (since it is pure S--wave,
Eq. (\ref{eq:sumdif})). Thus
\begin{eqnarray}
\aminusbn & = & \Frac{4 \alpha}{m_{\pi}} \, \lim_{s \rightarrow 0} \,
\Frac{f^N(s)}{s} \nonumber \\
\label{eq:swave}
& & \\
\aminusbc & = & \Frac{4 \alpha}{m_{\pi}} \, \lim_{s \rightarrow 0} \,
\Frac{[f^C(s) - f_{BORN}]}{s} \nonumber
\end{eqnarray}
with
\begin{equation}
f^N(s) = \Frac{2}{3} \, [ \, f_0(s) - f_2(s) \, ] \, \, \, \, \, \, \,
\, \, \, \, \, \, \, , \, \, \, \, \, \, \, \, \, \, \,
f^C(s) = \Frac{1}{3} \, [ \, 2 f_0(s) + f_2(s) \, ] \, \, \, \, \, \,
\, \, \, \, .
\end{equation}
The Omn\`es functions, $\Omega_I(s)$ in Eq. (\ref{eq:disp}), can be determined
from experimental phase shifts.
The structure of the left hand cut $p_I(s)$ and the subtraction constants
can be worked out from $QED$ at low energy and chiral dynamics:
$PCAC$ zeros \cite{PE92}, matching with $\chi P T$ \cite{DH93} or
$G \chi P T$ \cite{KM94}, respectively.
\par
The analysis of Morgan and Pennington \cite{MP91,PE92} inputs the
appearance of a near
threshold zero in the S--wave $\ggppcero$ amplitude from chiral dynamics
at $s=s_N$ to fix subtraction constants in Eq. (\ref{eq:disp}). The
remaining input information is taken from experiment. The S--wave
amplitudes for $\ggpp$ are thereby determined and so predictions
can be made for the neutral pion polarizability  $\aminusbn$. Not
surprisingly, given Eq. (\ref{eq:swave}), this combination is found
to be directly proportional to how far the chiral zero is from $s=0$.
(Note that at ${\cal O} (p^4)$ in $\chi$PT , this zero is at
$s_N = m_{\pi}^2$, while at ${\cal O} (p^6)$ is at
 \footnote{This can be seen in Fig. 8 of \cite{BG93}.}
$s_N \simeq 1.1 \, m_{\pi}^2$).
\par
Alternatively, others \cite{DH93,KM94}, rather than inputting
experimental information, have attempted to match these dispersion
relations with detailed predictions of $\chi$PT (to a given order) for
a range of energies about some point. Donoghue and Holstein \cite{DH93}
use $s_{\circ}=0$, while Knecht et al. \cite{KM94} try a range from
$s_{\circ}=0$
to $s_{\circ}= 4 m_{\pi}^2$. Knecht et al. find that the resulting
neutral pion polarizability depends strongly on the subtraction
point chosen for this matching \footnote{In doing this Knecht et al.
assume the combination $\aplusbn$ is saturated by vector and C--odd
axial vector mesons.}. Indeed, even the sign of the combination
of neutral pion polarizabilities, Eq. (\ref{eq:swave}), changes as
$s_{\circ}$ is varied. This implies that the $\chi$PT predictions
for $0 < s < 4 m_{\pi}^2$ must have a significantly different dependence
on energy than given by the other inputs to Eq. (\ref{eq:disp}). This
problem requires further work to isolate the source of the discrepancy.
We note that the dependence is far softer for the charged pion
polarizabilities.
\par
In these approaches, predictions are not just made for the polarizabilities
but for the amplitudes for $\ggpp$ scattering at low energies too.
Pennington
\cite{PE92} illustrates that an equally good description of low energy
data is obtained whether the chiral zero is at $s_N = \Frac{1}{2} m_{\pi}^2$
or $2 m_{\pi}^2$, not just $m_{\pi}^2$. However, the combination
$\aminusbn$ varies by a factor $4$ in this range of $s_N$.
Similarly, the ${\cal O} (p^6)$ $\chi$PT prediction of \cite{BG93} and the
${\cal O} (p^5)$ G$\chi$PT prediction of \cite{KM94} describe low energy
$\ggppcero$ data equally well, yet can have $\aminusbn$ with opposite
signs.

$$
\begin{array}{|c|c|c|c|}
\multicolumn{4}{l}{{\bf Table \; \; III} \; \; Results \; \; of \; \;
fits   \; \; to  \; \; the \; \;
experimental \; \; data \; \; on \; \; \ggpp \; \; by } \\
\multicolumn{4}{r}{\; \; \; \; \; \; \; \; \; \; \; \; \; \; \; \;
 \;  \; \; Kaloshin \; \; et \; \; al. \; \cite{KS94,KP94}
\; to \; \;
determine  \; \; the \; \; pion \; \; polarizabilities}
\vspace{0.5cm} \\
\hline
\hline
&
\multicolumn{3}{|c|}{}  \\
&
\multicolumn{3}{|c|}{Data \; \; fitted} \\
&
\multicolumn{3}{|c|}{}  \\
\cline{2-4}
& & & \\
\multicolumn{1}{|c}{Polarizability} &
\multicolumn{1}{|c}{CELLO} &
\multicolumn{1}{|c}{MARK-II} &
\multicolumn{1}{|c|}{Crystal \, \,  Ball} \\
& & &  \\
\hline
\hline
& & & \\
\aplusbc & 0.30 \pm 0.04^{*} & 0.22 \pm 0.06^{*}  & -  \\
& & & \\
\hline
& & & \\
\aminusbc & 5.3 \pm 1.0 & - & - \\
& & & \\
\hline
\hline
& & & \\
\aplusbn & - & - & 1.00 \pm 0.05^{*} \\
& & & \\
\hline
& & & \\
\aminusbn & - & - & -0.6 \pm 1.8 \\
& & & \\
\hline
\hline
\multicolumn{4}{l}{* \; \; See \; \; text \; \; for \; \; discussion \; \;
of \; \; these \; \; errors }
\end{array}
$$

In a related way Kaloshin et al. \cite{KS86,KS92,KS94} directly
parameterize  the dispersion relations of Eq.~(\ref{eq:disp}) in terms
of polarizabilities as the subtraction constants. Then by fitting the
charged and neutral channel data \cite{BO90,BE92,MA90} they find the
results for $(\alpha_{\pi} - \beta_{\pi})^{C,N}$ summarized in Table III.
Note the large uncertainty in the neutral combination reflects the
insensitivity of the low energy data to the exact position of the chiral
zero already mentioned. In \cite{KP94}, Kaloshin et al. have attempted
a similar analysis for the D--wave combination $(\alpha_{\pi} +
\beta_{\pi})^{C,N}$ and claim to have determined these with remarkable
precision  ---again their results are tabulated in Table III. In view
of the detailed amplitude analysis of the same data by \cite{MP90}
which shows the D--wave cross--section may be uncertain by as much as
$50 \, \%$ even in the $f_2(1270)$ region, the errors quoted by
Kaloshin et al. \cite{KP94} appear unbelievably small \cite{PE94}.
\newpage

\subsection{Quark Confinement Model (QCM)}
\hspace*{0.5cm}
The QCM developed by Efimov et al. \cite{IM92,EI93} has also been
employed to predict
values for the pion polarizabilities. The basic characteristic of the model
is a confinement ansatz that allows
loop calculations to be made finite, avoiding the need for a regularization
 procedure.
\par
Efimov et al. \cite{EI93} consider the contribution to pion polarizabilities
coming from pure loop terms and vector, axial vector and scalar resonances.
Their results are given in Table IV.
\par
As can be seen the most important contribution in this model is given
by the scalar resonances. Moreover, in the case of neutral pion
polarizabilities,
 there is a strong cancellation between loops and the scalar
contribution. The contribution of vector mesons is quite similar to that
 found in $\chi P T$, but not for axial vector mesons.
As mentioned earlier the main contribution to the leading order in $\chi P T$
to $\alphaplus$ comes from these axial--vector resonances, but in the
QCM scheme their
contribution is negligible.

\vspace {1cm}
$$
\begin{array}{|c|c|c|c|c|c|}
\multicolumn{6}{c}{{\bf Table \; \; IV} \; \; Predictions   \; \; of \; \;
QCM \; \; for \; \; the \; \; pion \; \; polarizabilities} \\
\multicolumn{6}{l}{\; \; \; \; \; \; \; \; \; \; \; \; \; \; \; \; \;
\; \; by \; \; Ivanov \; \; et \; \; al. \; \; \cite{IM92}}
\vspace{0.5cm} \\
\hline
\hline
& & & & & \\
\multicolumn{1}{|c}{Polarizability} &
\multicolumn{1}{|c}{Loops} &
\multicolumn{1}{|c}{0^{++}} &
\multicolumn{1}{|c}{1^{--}} &
\multicolumn{1}{|c}{1^{++}} &
\multicolumn{1}{|c|}{Total} \\
& & & & & \\
\hline
\hline
& & & & & \\
\alpha_{\pi^{\circ}} & -3.01 & 3.76  & 0  & 0  & 0.75 \\
& & & & & \\
\hline
& & & & & \\
\beta_{\pi^{\circ}}  &  2.95 & -3.76 & 0.51 & 0  & -0.30 \\
& & & & & \\
\hline
\hline
& & & & & \\
\alpha_{\pi^{\pm}}  & -0.14 & 3.76  & 0  & 0.02 & 3.64 \\
& & & & & \\
\hline
& & & & & \\
\beta_{\pi^{\pm}} & 0.30 & -3.76 & 0.05  & 0 & -3.41 \\
& & & & & \\
\hline
\hline
\end{array}
$$
\newpage

Let us comment on these predictions of the  Quark Confinement Model~:

\begin{itemize}
\item[1/] As stated before, the charged electric pion polarizability is
related
to the quotient of axial and vector structure--dependent form factors
of the radiative pion decay Eq. (\ref{eq:hav}). In the QCM $\alphaplus$ and
$h_A / h_V$ are independent quantities. How this model legitimately
avoids this chiral constraint is unclear.

\item[2/] There is a large model dependence in the scalar sector of QCM.
Moreover, the experimental parameters used in their model have changed
significantly \cite{PD92}. The authors themselves consider that the
contribution
of scalars has an error of at least $30 \%$. It could easily be bigger.
However, the scalar contributions cancel in $(\alpha +
\beta)$, giving
\begin{equation}
\aplusbc |_{QCM} = 0.22 \; \; \; \; \; \; \; \; \; \; \; , \; \; \; \; \;
\aplusbn |_{QCM} = 0.44 \, \, \, \, \, .
\end{equation}

\item[3/] Finally, there remains the question of double counting between loops
and meson resonances. This is an unresolved issue for models
that consider contributions at both the quark and meson level
simultaneously.

\end{itemize}

\subsection{QCD sum rules}
\hspace*{0.5cm}
Recently a calculation of the electric polarizability for the charged pion
has been performed using QCD sum rules \cite{LN94}. The authors take the
current algebra sum rule of Das, Mathur and Okubo \cite{DM67},
\begin{equation}
\alphaplus \, = \, \Frac{\alpha}{3 m_{\pi}} \langle r_{\pi}^2 \rangle \;
+ \; \Frac{\alpha}{2 m_{\pi} F_{\pi}^2} \; \int_{4 m_{\pi}^2}^{\infty}
\; d s \; \Frac{1}{s^2} \, [ \, \rho^A (s) \; - \; \rho^V (s) \, ] \; \; \; \;
\; \; ,
\label{eq:dmokubo}
\end{equation}
where the spectral functions are defined as in Eq.~(\ref{eq:espectro}).
They
input the experimental \cite{AM86} pion charge radius $\langle r_{\pi}^2
\rangle \, = \, (0.439 \pm 0.008) \, fm^2$ and then compute the integral
of the vector and axial spectral functions using QCD sum rule methods.
They obtain
\begin{equation}
\alphaplus \; |_{QCDsr} \; = 5.6 \pm 0.5 \; \; \; \; \; \; \; \; \; ,
\label{eq:qcdsr}
\end{equation}
the integral in Eq.~(\ref{eq:dmokubo}) largely cancelling the charge
radius term. Fortunately, Lavelle et al. find that their calculation
depends little on the poorly known quark condensate and so they deduce
a rather precise value for $\alphaplus$. This value, Eq.~(\ref{eq:qcdsr}),
is in good agreement with Eq.~(\ref{eq:serpukov}), but not with the
prediction of one loop $\chi$PT, Eq.~(\ref{eq:alp4}).

\section{Pion polarizabilities at DA$\Phi$NE}

\hspace{0.5cm} The DA$\Phi$NE electron--positron collider at $\sqrt{s} \,
\sim \, 1.02 \, {\rm GeV}$ provides the opportunity to study $\ggpp$ processes
at low energy. This is due to the possibility of double tagging of
$e^+ e^- \rightarrow e^+ e^- \pi \pi$ in the KLOE multi--particle
detector \cite{AL93} that will eliminate much of the background.
\par
Assuming full acceptance of the detector and with a machine luminosity
of $L \simeq 5 \times 10^{32} \; \rm{cm^{-2}} \rm{s^{-1}}$, the
expected rates are
\begin{itemize}
\item $N[\ggppcero ] \; \simeq \; 10^4 \, \, \rm{events/year}$
\item $N[\ggppplus ] \; \simeq \; 1.8 \times 10^6 \, \, \rm{events/year}$
\end{itemize}
Such event rates will allow the low energy cross--sections to be measured
with much higher statistics than previously.
\par
However, the pion polarizabilities are still likely to be poorly known.
The maximum $\pi \pi$ mass that can be realistically attained is $\sim 0.6 \;
\rm{GeV}$. This will allow the S--wave cross--sections to be accurately
measured,
but not the difference of the D--wave from its Born component without precise
azimuthal correlations. Even then, we have seen that knowing the low
energy $\gamma \gamma \rightarrow \pi \pi$ cross--section accurately still
allows large uncertainties in the polarizabilities \cite{PE94}. Only
measurements of Compton scattering will resolve these.
\par
In principle, there are two ways of studying Compton scattering on a pion.
Both involve $\gamma \pi$ production, initiated by either a photon or
a pion beam. The idea is that $\gamma N \rightarrow \gamma \pi N$ proceeds
by one pion exchange at small momentum transfers allowing Chew--Low
extrapolation to extract the Compton cross--section, while
$\pi Z \rightarrow \gamma \pi Z$ will occur by one photon exchange
if the scattering is on a heavy target, $Z$, and the photon is almost
real. Such a Primakoff production experiment, $E781$, is planned at
Fermilab for 1996 with an initial pion momentum of $600 \, {\rm GeV/c}$,
when the virtual photon can have a $4$-momentum squared of $-2 \times
10^{-8} \, {\rm GeV^2}$ very close to zero \cite{MO92,FE94}. The
process $\gamma N \rightarrow \gamma \pi N$ may prove possible with a
back--scattered laser experiment at an electron storage ring facility
such as Grenoble \cite{DA94,RE94}. These experiments are for the
future; they are much needed if we are to learn more of the pion's
polarizabilities.
\newpage
{\bf Acknowledgements}
\vspace*{0.4cm} \\
\hspace*{0.5cm}
This work has been supported in part by EC Human and Capital Mobility
programme EuroDA$\Phi$NE network under grant ERBCMRXCT 920026.

\end{document}